\documentclass[]{spie}  

\usepackage{amsmath,amsfonts,amssymb}
\usepackage{graphicx}
\usepackage{float}
\usepackage[colorlinks=true, allcolors=blue]{hyperref}

\title{The Role-Relevance Model for Enhanced Semantic Targeting in Unstructured Text}

\author{Christopher A. George}
\author{Onur Ozdemir}
\author{Connie Fournelle}
\author{Kendra E. Moore}
\affil{Boston Fusion Corp., 70 Westview Street, Suite 100, Lexington, MA 01824}
\authorinfo{Send correspondence to alex.george@bostonfusion.com \\ To appear in Proceedings of SPIE, Defense + Commercial Sensing: Next Generation Analyst (2018).}

\begin{document} 
\maketitle

\begin{abstract}
Personalized search provides a potentially powerful tool, however, it is limited due to the large number of roles that a person has: parent, employee, consumer, etc. We present the role-relevance algorithm: a search technique that favors search results relevant to the user's current role. The role-relevance algorithm uses three factors to score documents: (1) the number of keywords each document contains; (2) each document's geographic relevance to the user's role (if applicable); and (3) each document's topical relevance to the user's role (if applicable). Topical relevance is assessed using a novel extension to Latent Dirichlet Allocation (LDA) that allows standard LDA to score document relevance to user-defined topics. Overall results on a pre-labeled corpus show an average improvement in search precision of approximately 20\% compared to keyword search alone. 
\end{abstract}


\section{INTRODUCTION}
\label{sec:intro} 

Google's PageRank algorithm\cite{PageRank} has largely solved the challenge of document search on the internet. PageRank, however, is dependent on inter-document hyperlinks, and does not address the challenge of static documents filled with natural-language text. An orthogonal, but related, problem is how to filter results for the targeted end-user: for example, the ideal results from a search for ``Pythagoras'' will vary depending on whether the searcher is a philosophy professor (who would likely want technical descriptions of Pythagoreanism), a philosophy student (more likely to want an overview of the key ideas), or a middle schooler (probably more interested in examples of using the Pythagorean Theorem). Note that the desired results are attached to the role, not to the person -- a philosophy professor may require a refresher on the Pythagorean Theorem when acting as parent rather than professor.

We address this filtering challenge using the key idea of role. We first define a procedure to define roles and use entities and topics to label documents with their relevance to a given role. We then incorporate these roles into a keyword search for enhanced semantic targeting. Such an approach will be particularly useful in a scheme with fixed roles (e.g., a newsroom) or for searching a particular corpus (e.g., an encyclopedia). 

The main contributions of this work include (1) introducing and demonstrating role-based-search as a better-motivated alternative to personalized search, (2) presenting an extension to Latent Dirichlet Allocation (defined in Section \ref{sec:LDA}) as a way to determine documents' relevance to \textit{pre-defined} topics (which can be associated with roles), and  (3) demonstrating how simple knowledge structures can dramatically improve search precision.

\section{RELATED WORK}

Personalized search \cite{PersonalizedSearch} is a very similar concept to role-based search, in that it attempts to predict which articles a person will find relevant based on previous activity. We postulate, however, that personalized search is hamstrung by a user's many, dynamic roles -- professionals do not want cooking advice at work, nor do students want results optimized for courses they have completed. Further, modern examples of personalized search topics tend to have rather broad categories, leading to humorous results such as doctors receiving articles about how to ``get a career in healthcare and make \$50K per year!'' with the explanation ``you've shown interest in medicine.'' Further, the methods for learning the users' interests, such as scanning their e-mail, raise privacy concerns.

Collaborative filtering \cite{CF} is an alternative method in which people are clustered together based on their interests, and relevance judgments are propogated across the cluster. Though this has proven promising for recommender services such as Netflix, its applicability to document retrieval is limited because (1) it is difficult to get feedback from the user (asking the user ``was this helpful?'' for every article is obnoxious, but assuming that every opened article was helpful is highly error-prone), and (2) in a large corpus, many documents may only be read a few times, making it difficult to gather statistics. 

Natural language processing (NLP) is showing an increased capability to understand unstructured text through steps such as entity extraction \cite{entity}, event detection \cite{event}, and sentiment analysis \cite{sentiment}. This has proven useful for topics such as question-answering and populating knowledge bases; however, its applicability to document retrieval has thus far been limited. 

\section{DATASET}
\label{Reuters}

All numerical examples and results in this work were obtained using a collection of Reuters news articles currently administered by NIST \cite{Reuters}; the full dataset has over 2.6 million English-language articles.  In conjunction with this dataset, we use the 100 sample queries and relevance judgments from the 2002 Text REtrieval Conference (TREC) filtering challenge \cite{TREC}; these are useful as ground-truth for evaluation purposes.

\section{ELEMENTS OF ROLE-BASED SEARCH}

We now turn to the mechanics of implementing role-based search. In particular, we consider three ways to perform document filtering: using keywords, using entities, and using topics. We develop these separately without any notion of role; then, we combine them into a unified search engine and add the idea of role. 
\subsection{Keyword Search}

We use the Query Likelihood Model (QLM) \cite{QLM} for our keyword search. The query likelihood model calculates the relevance of each document to the search query according to the following equation:
\begin{equation} \textrm{score} = \textrm{nKeywordInDoc} + \frac{\mu \cdot \textrm{nKeywordInCorpus}}{\textrm{nTokens}} \label{QLM} \end{equation}
where  $\mu$  is a hyperparameter (we set it to 1000), nTokens is the number of vocabulary words (including duplicates) in the corpus, and the nKeywordInDoc and nKeywordInCorpus variables represent the number of times the keyword appears in the document and in the corpus, respectively. When multiple terms are searched, the scores for each term are multiplied (note, if N-gram extraction is performed, each N-gram is treated as a single search term). 
In Equation \ref{QLM}, the first term's relevance is obvious; the second term exists to smooth the corpus's response to rare words: a document missing a rare word is less indicative of irrelevance than a document missing a common word.
\subsection{Entity-based Search}
\begin{figure}[H]
\begin{center}
\includegraphics[width=11cm]{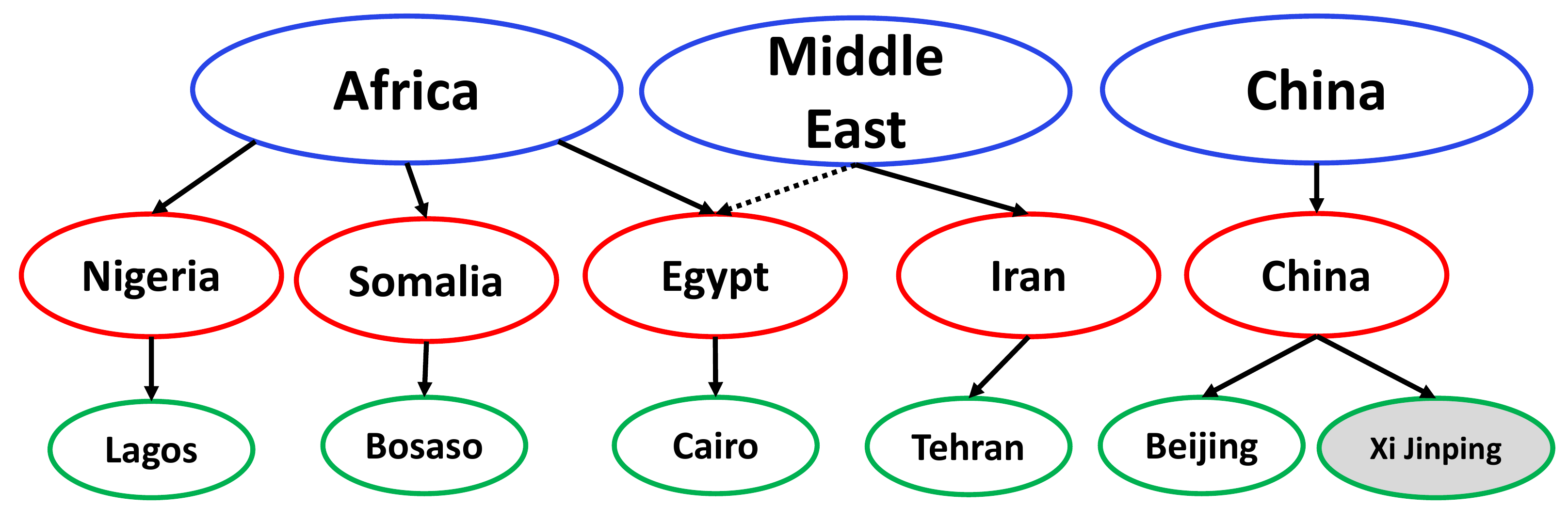}
\end{center}
\caption{Sample Geography Knowledge Structure (the dashed line illustrates that the weights need not be binary; the shaded node is an example of a person with geographical relevance)}
\label{fig:knowStruct} 
\end{figure}

Keywords offer search precision, returning documents that literally contain the exact term requested. In contrast, our other two filtering techniques attempt to mark documents as relevant or irrelevant to broad classes of queries. Our first approach is to leverage hierarchical knowledge structures, such as that illustrated in Figure \ref{fig:knowStruct}. 
\label{entities}

Knowledge structures can structure diverse types of knowledge, such as the subjects studied in academia or the areas of research at a hospital.  These structures could be useful for searching corpora of academic journal articles and medical records, respectively. In this work, we synthesized a knowledge structure using geographical locations. As illustrated in Figure \ref{fig:knowStruct}, our knowledge structure consists of three layers: a layer of global regions (top), a layer of countries, including disputed and former countries (middle), and a layer of cities (population $\geq$ 30,000) and people associated with each country (bottom). The cities are taken from GeoNames.com \cite{GN}.

We then use these entities to label each document with its relevance to the different layers of the knowledge structure. In some cases, the documents may come pre-labeled with entities: for example, the Reuters corpus (discussed in Section \ref{Reuters}) labels each article with a unique country. In other cases, it may be necessary to manually track which entities appear in each document (the cities, countries, and regions). We then calculate the document's distribution of entities over regions and over countries: for example, an article that mentions Beijing three times and Tehran once will be recorded as 75\% relevance to China and 25\% relevance to the Middle East and to Iran. When the searcher indicates interest in a given entity (e.g., to the Middle East), we then rank the documents by their relevance to that entity. The advantage is that a document that mentions Tehran will be marked as relevant to the Middle East, even if the phrase ``Middle East'' never appears in the text.

This entity search can be invoked whenever the user types a keyword that that appears in one of the  knowledge structures.  Alternatively, we can use entity-based search as part of a role, as discussed in Section \ref{roles}.
\subsection{Topic-based Search}
\subsubsection{Latent Dirichlet Allocation}
\label{sec:LDA}

In some cases, a knowledge structure may not be practical: for example, the different topics in the corpus may be very broad, interconnected, or technical, requiring knowledge structures that are large, expensive to maintain, unwieldy, or require substantial domain expertise. In these cases, we extract the latent topic distribution from the corpus using Latent Dirichlet Allocation (LDA) \cite{LDA}.

\begin{figure}[H]
\begin{center}
\includegraphics[width=12cm]{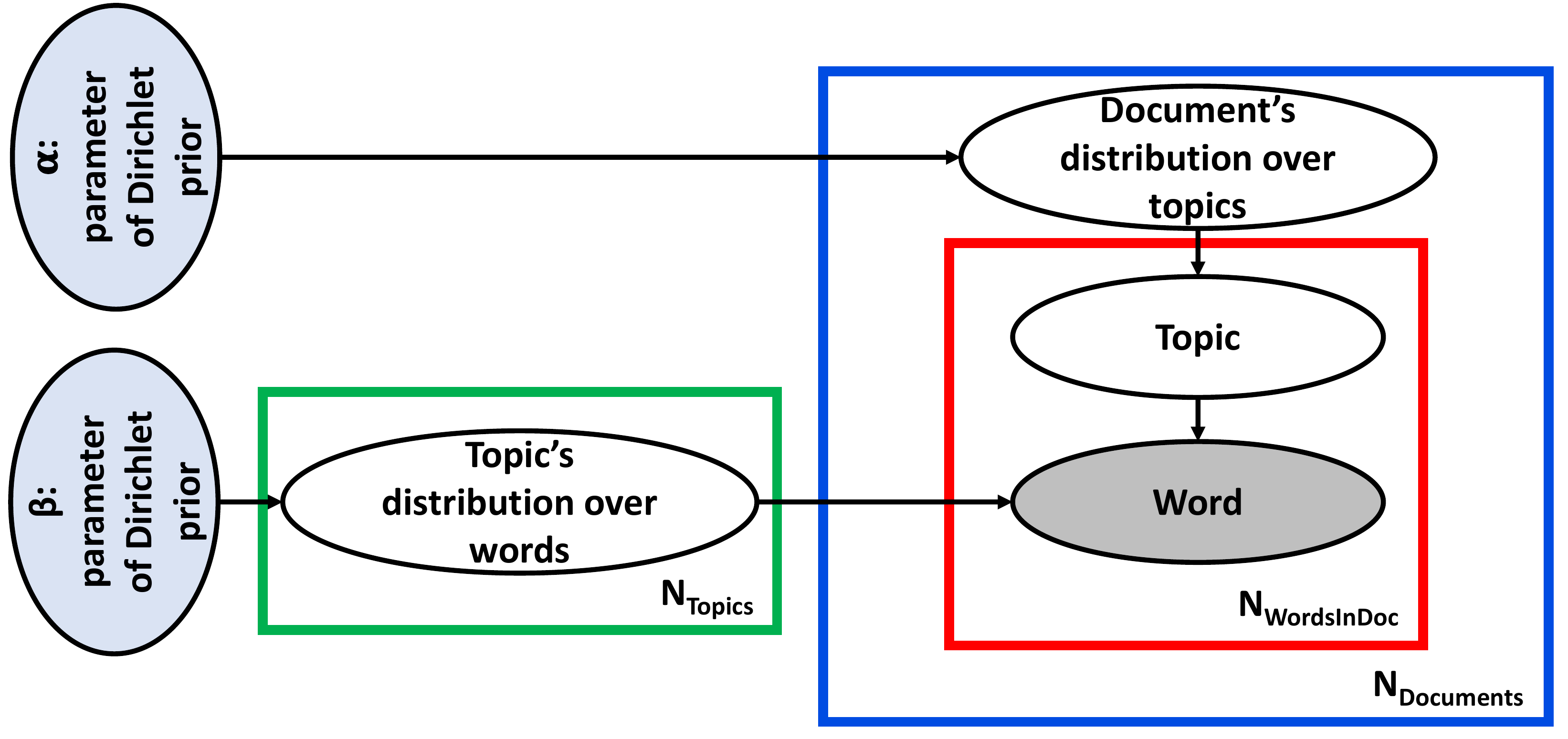}
\end{center}
\caption{Latent Dirichlet Allocation in Plate Notation}
\label{fig:example} 
\end{figure}

We do not discuss the various assumptions and limitations of LDA here; instead, we give a brief overview of the technique and demonstrate its efficacy. 
Figure \ref{fig:example} shows the probabilistic graphical model for LDA in plate notation. 
As Figure \ref{fig:example} shows, LDA is a generative model that postulates that the words on the page (shaded gray) are generated stochastically from a series of underlying latent distributions. 
In our case, we have the inverse problem: we know the words on the page, and we know that the left-most distributions are Dirichlet distributions (the values of $\alpha$ and $\beta$ are not known, but we can choose them by trial and error). 
In principle, we can use Bayesian inference to calculate the other latent distributions. 
In practice, attempting to solve the exact Bayesian inference leads to intractable integrals, so we approximate the topic distributions using a sampling technique such as Markov Chain Monte Carlo or Gibbs Sampling. 

Gibbs Sampling loops over the tokens (each instance of each word in the vocabulary) and begins by randomly assigning each token to a ``topic,'' where the user chooses the number of topics ($N$).  On subsequent passes through the tokens, the $i$th token is assigned to the $j$th topic with a probability given by: 
\begin{equation} P(z_i = j) = \frac{\textrm{(W-T count)} + \beta}{\textrm{nWordsAssToTopic} + W\beta} \cdot \frac{\textrm{(D-T count)} + \alpha}{\textrm{nWordsInDoc} + T\alpha} \label{gibbs} \end{equation}

The result is a matrix showing the number of times a given word was assigned to each topic (the W-T count); a separate matrix tracks the number of times any word in a given document was assigned to each topic (the D-T count). This allows each word and each document to be represented as a distribution over the topics; further, it allows each topic to be represented by the words most commonly assigned to it. 

To demonstrate the efficacy of LDA, we ran it on a real-world collection of news articles. Table \ref{tab:LDAstandard} shows the 7 words assigned most often to the first eight categories; these topics are manifestly coherent.

\begin{table}[h]
\centering
\begin{tabular}{|c|c|c|c|c|c|c|c|}
\hline
Topic 1 & Topic 2 & Topic 3 & Topic 4 & Topic 5 & Topic 6 & Topic 7 & Topic 8 \\ 
\hline
dividend           &   percent   &   tonne   &   trade      &    pope    &    drug        &    company    &    israel     \\
\hline
bank               &   growth    &   vessel  &    wto       &   visit    &   patient      &   share       &    israeli    \\
\hline 
industy            &   rate      &  port     & united state &  first     &   health       &   ltd         & palestinian   \\
\hline
bond               &  percent    &  ship     & mexico       &  church    &  treatment     &   firm        & netanyahu     \\
\hline
agm                &  year       & wheat     & agreement    & vatican    &  medical       &    deal       & arab          \\
\hline
half yearly result &  market     & sea       & chile        & day        & disease        & investment    & jerusalem     \\
\hline
finance            &  demand     & boat      & talk         & born       & cancer         & bid           & arafat        \\
\hline
\end{tabular}
\label{tab:LDAstandard}
\caption{Topics extracted by LDA in Reuters Corpus}
\end{table}

\subsubsection{Extending LDA to determine document relevance to pre-defined topics}

The challenge with using LDA for search is that it is an unsupervised approach: though it extracts many good topics, those topics the searcher is interested in will not necessarily align with any of the topics discovered. One way around this is to calculate the distance (e.g., cosine or Mahalanobis) between the query word's word-topic vector and the word-topic vectors of the words in each document; however, this is computationally expensive; further, it may be frustrated by words with multiple meanings.

\begin{figure}[H]
\begin{center}
\includegraphics[width=12cm]{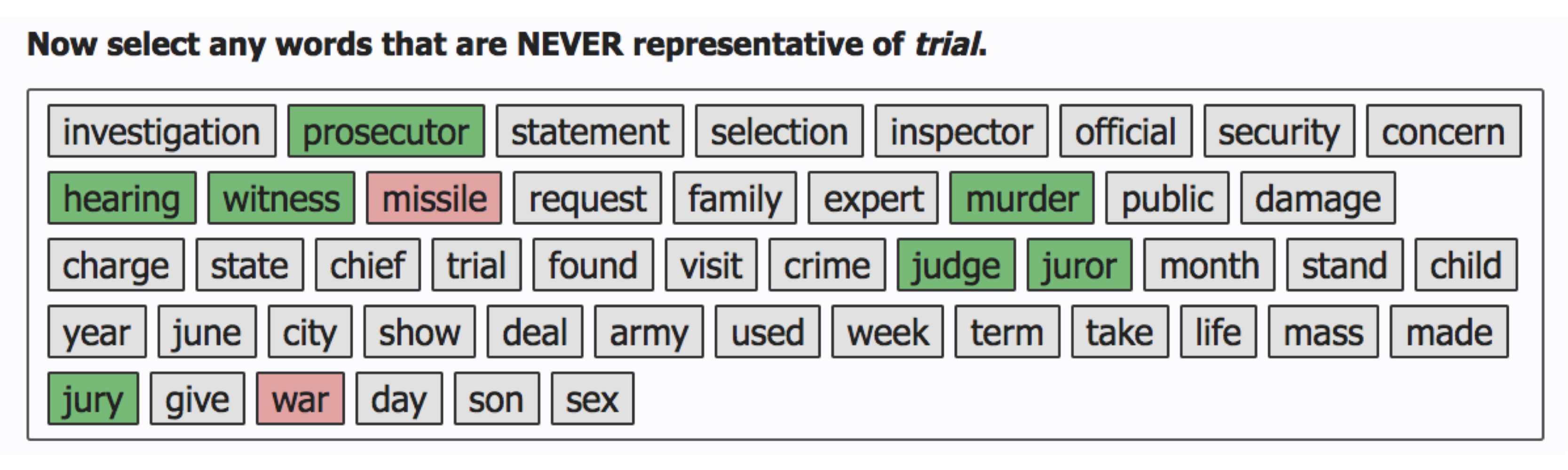}
\end{center}
\caption{Sample Topic Definition UI}
\label{fig:UI} 
\end{figure}

Our approach is instead to define the topics of interest -- this can be done once when the corpus is initialized, and need not be done at runtime by the end user. These topics are defined by entering keywords; specifically, the initializer enters a keyword to indicate their topics of interest (e.g., ``natural disasters''); we then find words with similar topic distributions and ask them to indicate which are relevant or irrelevant (Figure \ref{fig:UI} demonstrates a  UI for the user to make their selections). Note, this is not simply searching for synonyms of the keywords, but searching for related words -- for example, given a ``sports'' query, one would select words such as ``basketball,'' ``soccer'', ``football,'' etc.

We now use these keywords to represent the pre-defined topics as a linear combination of the unsupervised topics extracted by LDA. In principle, we can do this by averaging the topic vectors of the keywords directly, but these words' topic vectors are often known only coarsely due to the corresponding words appearing relatively rarely in the corpus. Therefore, we instead identify documents that are clear hits to the selected keywords and average the topic distributions for these documents (optionally, additional tuning can be performed to optimize results).  

Having now expressed both the user-specified topic and the documents as distributions over the LDA-extracted topics, it is straightforward to calculate the distance between the user-defined topic and the documents, thereby judging each document's relevance to the user-specified topic. In this way, we can use LDA even in a supervised setting where the desired topics are pre-defined (alas, we cannot call this procedure ``supervised LDA,'' as this term has already been claimed by the solution to a separate, though related, challenge \cite{sLDA}). 

Table \ref{tab:LDAresults} shows selected results from our labeling procedure: we show the titles of 15 selected documents and the corrected distance between the user-specified topic's distribution and each document's topic distribution For each topic, we automatically identified articles that had distances indicating they were on the border of relevant and irrelevant, and asked the user for relevance judgments for common words in the titles of those articles. Based on those judgments, we applied a correction to the distance values; this correction is responsible for the negative values in the table. In the first five rows, we show documents that are ranked as highly relevant to the ``natural disasters'' query. Indeed, the titles describe earthquakes and floods. The next five articles shown have higher distances from the query and therefore show reduced relevance; indeed, the titles refer to accidents, crimes, and conflicts, which are not natural disasters but do have many similar qualities. The final selected set of five articles have very high distances from ``natural disasters'', and represent articles that are not relevant to the query at all.

\begin{table}[h]
\centering
\begin{tabular}{|r|c|}
\hline
Corr. distance & Document Title \\
\hline
\hline
-0.3017 & PAKISTAN: Quakes in Iran, Pakistan kill 150. \\
\hline
-0.1942 & IRAN: More than 550 die in three Asian quakes. \\
\hline
-0.1413 & PAKISTAN: Nearly 50 killed in Pakistan earthquake. \\
\hline
-0.1276 & IRAN: Over 400 killed in three Asian quakes.  \\
\hline
-0.1225 & ICELAND: FEATURE - Iceland glacier flood is unique. \\
\hline
\hline
 0.2413 &  USA: Fire aboard Cunard cruise liner, one dead. \\
\hline
 0.2414 &  INDIA: Bus and taxi bombs kill one, injure 11 in Delhi.  \\
\hline
 0.2416 &  SRI LANKA: Sri Lanka seizes rebel arms ship, sinks boat. \\
\hline
 0.2417 &  ALBANIA: Albania shows few signs of renewal amid collapse. \\
\hline
 0.2420 &  USA: Casino trip bus hijacked, driver shot in face. \\
\hline
\hline
 0.4805 &  ESTONIA: Estonian bus crash kills five children. \\ 
\hline
 0.4805 &  USA: Viacom ends talks with MCA over TV venture. \\
\hline
 0.4805 &  AUSTRALIA: RTRS-Australian Broadcasting Corp Morning Update. \\
\hline
 0.4805 &  MEXICO: Opposition ahead in Mexico City governor race - poll. \\
\hline
 0.4805 &  USA: U.S. Sen. Specter has benign brain tumor removed.  \\
\hline
\hline
\end{tabular}
\caption{Document Relevance Scores for ``Natural Disaster'' in Reuters Corpus (smaller values indicate more relevant documents.)}
\label{tab:LDAresults}
\end{table}

\subsection{Multi-pronged and Role-Based Search}
We now combine these three search methodologies into a single search decision. There are clear advantages to this multi-pronged approach.
\label{roles}
\begin{itemize}
\item The keyword search offers specificity: it is the only search that focuses its results specifically on the user input, rather than on broad categories such as topics or branches of a knowledge structure. This approach requires the exact keyword to appear in the document; for example, a keyword search for the word ``sports'' will return many articles about sport utility vehicles, as these articles explicitly use the word ``sport'' many times. Conversely, articles about basketball or soccer may never explicitly contain the word ``sport'', and so will not be returned. Depending on the desired use-case, this may not be the desired behavior.
\item The topic-based search ranks the relevance of the document as a whole, rather than simply the presence or absence of a particular word.  This methodology can be expected to correctly identify those documents related to ``sports'', even if the word ``sport'' never appears. Due to the inherently statistical approach and the mechanics of LDA, however, topic-based search works best for broad topics that contain a large number of articles; therefore, topic-based search does not return the precision of a keyword search. Further, defining topics is (in this work) performed manually, which puts a practical limit on the number of topics that can be defined.
\item The entity-based search, like the topic search, identifies a document's relevance to a particular topic by tracking the entities that occur in the document and comparing them to a knowledge structure. This offers the key advantage that it can leverage the entire knowledge base rather than just the keyword (e.g., the knowledge base depicted in Figure \ref{fig:knowStruct} ``knows'' that Tehran is in Iran and therefore in the Middle East); however, it requires that such a knowledge base exist. 
\end{itemize}

These three approaches therefore complement each other. We calculate the score from each methodology separately: the topic and entity scores are between 0 (not relevant) and 1 (maximally relevant); we convert these to a z-score (i.e., we express them as a number of standard deviations above or below the mean). The keyword (QLM) score is an absolute value: we do not make this a z-score, as a given query may have no relevant results or several highly-relevant results. We then combine these as a convex combination, as given by:
\begin{equation} \textrm{score} = \lambda_1 \cdot \textrm{TopicZscore} + \lambda_2 \cdot \textrm{EntityZscore} + (1-\lambda_1-\lambda_2) \cdot \textrm{QLMscore} 	\label{eqn:role} \end{equation}
We empirically choose $\lambda_1$ = 0.07 and $\lambda_2$ = 0.90 for our tests.

We now add the idea of role. For example, consider a role such as an economic analyst for domestic affairs: this would entail both a topic (``economics'') and a geographical region (``USA''). In principle, we could simply append ``USA'' and ``economics'' to every search term this analyst performs -- however, this would simply search for those additional words, without any knowledge of the constituents of the USA, nor of concepts related to economics. Instead, we run the corresponding entity- and topic-based searches, and append these queries to the analyst's keyword search according to equation \eqref{eqn:role}.  This ensures that all keyword searches will preferentially return results relevant to the user's role. For example a search for ``potato'' will return both (a) the highest-scoring articles for the keyword ``potato'' and (b) lower-scoring articles that mention the keyword ``potato'' and also are particularly relevant to economics and/or the USA. 

\section{EXPERIMENTS}
In the previous section, we claimed that simply adding role-related search terms to the user's query should be far less effective than running the entity-based and topic-based searches. In this section we describe our numerical experiments that prove this claim, starting with a ``raw'' natural language corpus and ending with search evaluations.

\subsection{Extract, Transform, and Load}

\indent Our first step upon unpacking a corpus is to extract the unstructured text from each document (e.g., by using Beautiful Soup\cite{BS}) and to extract the titles from each article. 

We next extract a vocabulary from the unstructured documents. We begin by ``lemmatizing'' each word: plural nouns are returned to the singular form, conjugated verbs are returned to the stem, etc. Next, very common words such as articles, demonstratives, and prepositions, are culled. The remaining words are then counted and ranked by how often they appear in the corpus. We then create a vocabulary from these words:
\begin{itemize}
\item	The top 10,000 words are added to the core vocabulary. This vocabulary is used throughout the remainder of the Extract, Transform, and Load (ETL) pipeline and in the role definitions.
\item	The top 100,000 words are stored for the keyword search only. We find that this is a good balance between the competing demands of precision and performance.
\end{itemize}

The tokenization step then represents each article as a list of numbered vocabulary words. Table \ref{tab:tokens} gives an example; note that the very common words (``stop words'') are removed. 

\begin{table}[H]
\centering
\begin{tabular}{|c|p{12.5cm}|}
\hline
Original Text & ``NYCE November frozen concentrated orange juice futures closed higher in thin dealings as players dismissed an encroaching tropical storm in the Gulf of Mexico.'' \\
\hline
Tokenized Text & ``november frozen concentrated orange juice future closed higher thin dealing player dismissed tropical storm gulf mexico'' \\
\hline
Tokens & ``6092, 3659, 1852, 6238, 4814, 3693, 1654, 4154, 9024, 2303, 6655, 2631, 9270, 8589, 3976, 5608'' \\
\hline
\end{tabular}
\label{tab:tokens}
\caption{Example of Extracted Text}
\end{table}

After tokenization, we extract ``phrases'' (N-grams), which are common sequences of words. To extract the phrases, we use the following standard procedure\cite{PM}:
\begin{itemize}
\item Pass through the articles and record every possible bigram and trigram (phrases of length two and three, respectively). Obtain the count of each phrase.
\item Determine the significance of each bigram using equation \ref{bigrams}, where nPhrase is the count from step 1, nWordN is the number of occurrences of the $N$th word in the corpus, and nTokens is the number of words (including duplicates) in the corpus. Note that equation \ref{bigrams}  sets the score to zero for words that are distributed uniformly throughout the corpus; this prevents the most common words from forming insignificant phrases with each other.
\begin{equation} \textrm{score} = \textrm{nPhrase} - \frac{(\textrm{nWord1})(\textrm{nWord2})}{\textrm{nTokens}}	\label{bigrams} \end{equation}
For trigrams, the reasoning in the same; the score is given by equation \ref{trigrams}:
\begin{equation} \textrm{score} = \textrm{nPhrase} - \frac{(\textrm{nWord1})(\textrm{nWord2})(\textrm{nWord3})}{(\textrm{nTokens})^2} 	\label{trigrams} \end{equation}
\end{itemize}
	The bigrams and trigrams with the highest score are kept. We typically require the number of phrases to equal 15\% of the number of words. 
Some examples of extracted phrases are shown in Table 4.

\begin{table}[H]
\centering
\begin{tabular}{|c|c|c|c|c|}
\hline
prime minister & last year & hong kong & last week & year old \\
\hline
year ago & told reporter & european union & news conference & per cent \\
\hline
west bank & last month & white house & five years & south korea \\
\hline
\end{tabular}
\label{fig:phrases}
\caption{Example of Extracted Phrases}
\end{table}

\subsection{Role Definitions}
Having extracted the tokens, we now define the roles on our Reuters corpus of news articles:
\begin{itemize}
\item We use the entity-based search for geography as in Section  \ref{entities}. In particular, we define 11 global regions including the USA, Russia, China, the Middle East, etc.; we then create a knowledge structure that contains all the countries and cities in each region. (Note, we remove a small number of cities whose names overlap with common nouns, such as Nice and Post). 
\item We use the topic-based search for news-relevant topics such as politics, economics, religion, health, law, disasters, and sports. 
\end{itemize}

\subsection{Evaluation Results}
We use the TREC relevance judgments (discussed in Section \ref{Reuters}) to quantify the advantage of role-based search. 

We begin with a strong positive example: Table \ref{tab:STARTERres2} shows the results for a task to ``find articles related to the impact of terrorism on tourism in the Middle East.'' As the table shows, searching for the keywords ``terrorism'' and ``tourism'' gives only one relevant result out of the first 20 results returned. Adding ``Middle East'' to the query does not help at all, likely because relevant articles discuss specific countries in the Middle East, not the region as a whole. Switching ``Middle East'' from a keyword to an entity increases the search precision by a factor of 3. We then switch ``terrorism'' from a keyword to a topic; this increases the search precision by an additional factor of five, likely because the word ``terrorism'' is too general (an article about suicide bombings might never use the word terrorism, but it is easy to identify terrorism-relevant documents from the topic distribution). Thus, setting role in this case increased search precision by a factor of 15.

\begin{table}[h]
\centering
\begin{tabular}{|c|c|}
\hline
Method & Number relevant results (out of 20) \\
\hline
Keyword: ``Tourism Terrorism'' & 1 \\
\hline
Keyword: ``Tourism Terrorism Middle East'' & 1 \\
\hline
Keyword: ``Tourism Terrorism'' + Entity: Middle East & 3 \\
\hline
Keyword: ``Tourism'' + Entity: Middle East + Topic: Terrorism & 15 \\
\hline
\end{tabular}
\caption{Example Evaluation Result using Reuters Corpus and TREC 2002 Relevance Judgments \label{tab:STARTERres2}}
\end{table}

Table \ref{tab:STARTERres1} gives a more systematic result by showing the average number of relevant results appearing in the top-20 results for various search methodologies over several queries. As the table shows, adding the location as an additional keyword does increase the search precision significantly, but it is far more effective to add the location using entity-based search. This is logical as entity-based search has access to every city and country in the specified region, whereas keyword search merely searches for the name of the country in the article.

\begin{table}[h]
\centering
\begin{tabular}{|c|c|}
\hline
Method & Number relevant results (out of 20) \\
\hline
Keyword only & 6.7 \\
\hline
Keyword + Location as add'l keyword & 10 \\
\hline
Keyword + Location via entity relevance  & 12.1 \\
\hline
\end{tabular}
\label{tab:STARTERres1}
\caption{Evaluation Results using Reuters Corpus and TREC 2002 Relevance Judgments \label{tab:STARTERres1}}
\end{table}

Table \ref{tab:STARTERres1} proves that entity-based search is approximately 20\% more effective than simply adding the entity as an additional keyword. Table \ref{tab:STARTERres2} strongly suggests that the adding topic-based search further enhances search precision.

\section{EXTENSIONS}
\subsection{Multilingual Role-Relevance}

We now consider adapting the above approach to corpora containing multilingual documents, in which the goal is to type in English words and retrieve relevant documents regardless of language. The simplest approach is simply to translate the documents to English; even imperfect machine translations do not pose a significant problem since our tools care about the tokens in each article, not the ordering. Assuming that this is not an option, however, we consider what can be done given only a machine-parsable translation dictionary between English and the target language(s) for each of our three search methodologies: 
\begin{itemize}
\item Keyword search. Extending keyword search is straightforward given the translation dictionary: we simply use the dictionary to translate the keyword into the foreign language. When multiple translations are possible, we use the translation used most often in the foreign language corpus. This can, however, lead to some sub-optimality: for example, ``president'' is translated in German to ``Generaldirektor'', which is appropriate for a commercial context but not a political one. Such problems are common in machine translation and largely intractable, especially when translating single words, due to ambiguities in the source language, the non-injective nature of linguistic pairings, and cross-cultural differences. Despite this sub-optimality, this works reasonably well -- further, it could be improved by assigning the most appropriate translation(s) for words for each given role. 
\item Entity-based search. This is straightforward except that the list of entities (e.g., of cities and countries)  must be translated into the appropriate language. Given that this must be done only once per language and for only approx. 13,000 cities, we simply assume that this list can be translated or an alternative list found in the target language.
\item Topic-based search. This is relatively difficult. LDA on a multilingual corpus will result in topics of each language being discovered separately: for example, there might be a ``terrorism'' topic in English and a ``Terrorismus'' topic in German, but no way to match them up. We could run LDA on each language separately, but then our topic-based search will only ever recommend English-language documents. Even if we require the user to run LDA once per language, it is then unclear how to normalize. Such challenges can be avoided by replacing LDA with a multilingual analog such as Boyd-Graber and Blei's Multilingual Topic Model (MuTo) \cite{MuTo}. Given the topic distributions, we can run topic-based search as usual. 
\end{itemize}
We conclude that our role-relevance model is extendable to multilingual search, and the bottleneck is in the multilingual extension to LDA. Indeed, we ran some experiments with a bilingual English-German corpus; it consistently returned relevant German articles in response to English queries, though there was a clear bias toward English-language articles. Further study in this area is warranted.

\subsection{Streaming Documents}
In order to support documents that are not available when the corpus is initialized but become available during runtime, we must update the necessary parameters (e.g., the document-role matrix) when new documents are detected. It is straightforward to update most of the parameters on-the-fly: for example, we count the vocabulary words and entities in new documents and estimate their topic distributions based on pre-defined topics. Topics and vocabulary words may drift over time, particularly if the streaming documents are not similar to the original documents, and so the vocabulary words and topic distributions will need to be periodically recalculated. 

\section{CONCLUSIONS}
In this work, we proposed role-based search as a better-motivated alternative to personalized search. We defined roles using two types of filters: entity-based filters, which searches for entities based on their hierarchical relations in a knowledge structure, and topic-based filters, which use the LDA-derived topic distributions to relate documents to a user-defined topic of interest. These types of filters use the entire semantic content of the document to make relevance determinations; in combination with the precision of a keyword search, this multi-pronged approach offers a powerful search capability. Further, the topic and entity filters can be defined as a role, allowing the user to automatically find documnets relevant to their role even when searching only by keywords.

Our focus in this paper is on the entity, topic, and keyword searches discussed above. In future work, we will explore additional types of filters; for example, we could filter results by ``level'' (distinguishing beginner-level documents from expert-level documents) and document quality. We can also investigate tracking the user behavior and learning patterns for each role; this could potentially encode institutional memory in the role definitons (though it is difficult to assess which documents are useful to the user without constantly asking for feedback). We could also expand our software package to present a practical, perhaps open-source, search capability, and continue studying the extension to multilingual corpora.

\acknowledgments 
This work was funded by the Office of Naval Research under contract N00014-14-C-0245.

\bibliography{main} 
\bibliographystyle{spiebib} 

\end{document}